\documentclass[a4paper,twocolumn,citeautoscript,prl]{revtex4-1}

\usepackage{amsmath}
\usepackage[dvips]{graphicx,color}

\setcitestyle{super}

\begin{document}

\title{Observation of Arnold Tongues in Coupled Soliton Kerr Frequency Combs}

\author{Jae K. Jang$^{1}$}
\author{Xingchen Ji$^{2,3}$}
\author{Chaitanya Joshi$^{1,4}$}
\author{Yoshitomo Okawachi$^{1}$}
\author{\\Michal Lipson$^{1,3}$}
\author{Alexander L. Gaeta$^{1,3,*}$}

\affiliation{$^{1}$Department of Applied Physics and Applied Mathematics, Columbia University, New York, NY 10027, USA
\\$^{2}$School~of~Electrical~and~Computer~Engineering,~Cornell~University,~Ithaca,~NY~14853,~USA
\\$^{3}$Department of Electrical Engineering, Columbia University, New York, NY 10027, USA
\\$^{4}$School of Applied and Engineering Physics, Cornell University, Ithaca, NY 14853, USA}

\begin{abstract}
  \noindent We demonstrate various regimes of synchronization in systems of two coupled cavity soliton-based Kerr frequency combs. We show sub-harmonic, harmonic and harmonic-ratio synchronization of coupled microresonators, and reveal their dynamics in the form of Arnold tongues, structures that are ubiquitous in nonlinear dynamical systems. Our experimental results are well corroborated by numerical simulations based on coupled Lugiato-Lefever equations. This study illustrates the newfound degree of flexibility in synchronizing Kerr combs across a wide range of comb spacings and could find applications in time and frequency metrology, spectroscopy, microwave photonics, optical communications, and astronomy.
\end{abstract}

\maketitle

\noindent Synchronization is a universal feature of networks of coupled nonlinear oscillators~\cite{strogatz_from_2000,pikovsky_synchronization_2003}. Observed in various disciplines of science including ecology, neuroscience, chemistry and physics, synchronization manifests itself as the spontaneous adjustment of the natural rhythms of the interacting oscillators to a common frequency, leading to their mutually phase-locked state. Despite the vast diversity in the underlying processes across these systems, the phenomenon of synchronization can be investigated within a unifying mathematical framework~\cite{winfree_biological_1967,kuramoto_self_1975}. The Kuramoto model~\cite{kuramoto_self_1975}, in particular, predicts the existence of a threshold coupling strength at which a nucleation of phase-locked oscillators forms and grows in a self-sustaining manner. In this sense, synchronization is a remarkable manifestation of self-organization in which an ordered state emerges via a phase transition~\cite{strogatz_from_2000}, and the understanding of this phenomenon can provide invaluable insight into the dynamics of complex processes. In addition to being of considerable interest in fundamental science, synchronization also finds a myriad of applications. Radio-controlled clocks are periodically corrected by a central reference clock to achieve precise time-keeping~\cite{pikovsky_synchronization_2003}. A mechanical ventilator and cardiac resynchronization therapy~(CRT) device entrain the respiratory~\cite{graves_respiratory_1986} and cardiac cycles~\cite{anishchenko_entrainment_2000} of a patient, respectively. One can implement tunable local oscillators operating at millimeter and sub-millimeter wavelengths by utilizing synchronized Josephson junction arrays~\cite{wiesenfeld_frequency_1998}. In the field of photonics, synchronization process has been harnessed in lasers~\cite{nixon_controlling_2012} to achieve phase-locked arrays, and high optical power by means of coherent beam combining~\cite{goldberg_injection_1985}. In addition, timing synchronization of the ultrashort pulse trains from two mode-locked lasers has been experimentally demonstrated~\cite{cundiff_colloquium_2003} and used for pump-probe studies~\cite{potma_high_2002}.

Optical frequency combs have become a prominent subfield of photonics that has resulted in breakthroughs across various fields in engineering and science~\cite{cundiff_colloquium_2003,udem_optical_2002,newbury_searching_2011}. Recently, nonlinear microresonators driven by continuous-wave~(CW) sources have emerged as a highly promising platform for the generation and control of optical frequency combs~\cite{kippenberg_microresonator_2011,moss_new_2013}. Such microresonator-based frequency combs, also known as Kerr combs, can exhibit mode-locking behavior~\cite{saha_modelocking_2013,herr_temporal_2014,jang_observation_2014,jang_temporal_2015,delhaye_phase_2015,
liang_high_2015,yi_soliton_2015,joshi_thermally_2016,webb_experimental_2016,yi_single_2017,bao_soliton_2017,
obrzud_temporal_2017,liao_pump_2017,kippenberg_dissipative_2018,he_self_2018,
gaeta_photonic_2019,lu_deterministic_2019} that is associated with the formation of temporal cavity solitons~\cite{leo_temporal_2010} circulating in the microresonator. Subsequent studies have revealed that more than one cavity soliton-based Kerr comb can be simultaneously sustained in a single device by employing multiple spatial mode families~\cite{lucas_spatial_2018} or two pumps operating in the counter-rotating configuration~\cite{yang_counter_2017,joshi_counter_2018}. The counter-rotating cavity solitons were shown to exhibit a locking range in which their repetition rates become equal. Recently, the ability to synchronize two soliton Kerr combs in physically distinct microresonators has been theoretically examined~\cite{wen_synchronization_2015,munns_novel_2017} and experimentally demonstrated~\cite{jang_synchronization_2018}. In this experimental study, repetition rate-locking of two cavity soliton combs in the master-slave configuration was demonstrated over a distance larger than 20~m, and the enhancement of output comb power via coherent combining was also shown. However, these previous studies were performed under the assumption that the coupled microresonators were similar in length such that the frequency spacings of the two uncoupled combs were nearly equal. Nevertheless, there are multiple instances in the scientific literature suggesting that the ratio of entrained frequencies need not be unity~\cite{matsumoto_chaos_1987,zeng_theoretical_1990,sacher_intensity_1992,simonet_locking_1994}, and sub-harmonic frequency locking of a single-mode semiconductor laser by the optical injection of an intensity modulated beam has recently been demonstrated~\cite{shortiss_harmonic_2019}.

Here, we experimentally demonstrate sub-harmonic, harmonic and harmonic-ratio synchronization of mode-locked Kerr frequency combs. We employ three silicon nitride ($\mathrm{Si}_3\mathrm{N}_4$) microresonators of different radii. By examining all possible permutations of two out of the three microresonators in turn, we investigate various orders of synchronization, including rational (non-integer) cases. More specifically, we show evidence of the $q:r$ synchronization regime such that $q\Omega_\mathrm{M} = r\Omega_\mathrm{S}$, where $q,r = 1,2,3$ are integers, and $\Omega_\mathrm{M}$ and $\Omega_\mathrm{S}$ are spacings of the master and slave combs, respectively. For the first time, we reveal features widely known as Arnold tongues~\cite{pikovsky_synchronization_2003,shortiss_harmonic_2019} in a parameter space for systems of coupled microresonators that illustrates the general applicability of our study for exploring complex real-life systems. These results significantly expand the scope of the Kerr comb synchronization and could have implications in optical telecommunications~\cite{levy_high_2012,marinpalomo_microresonator_2017}, spectroscopy~\cite{suh_microresonator_2016,yu_microresonator_2018,dutt_onchip_2018}, and astronomy~\cite{suh_searching_2019,obrzud_microphotonic_2019}.

\begin{figure}[b]
\centerline{\includegraphics[width=1\columnwidth]{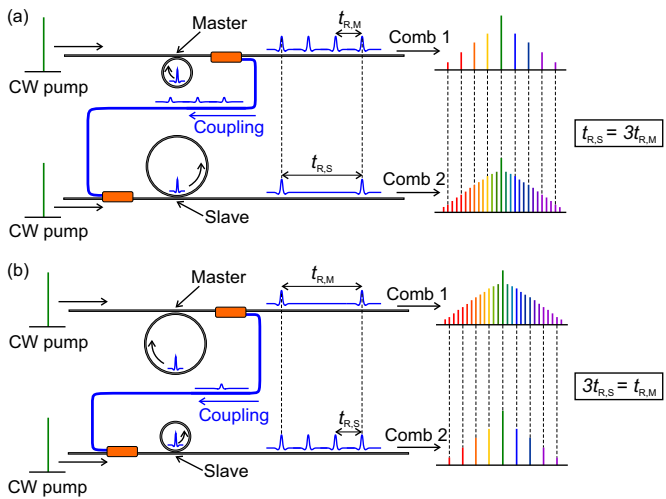}}
\caption{Schematic illustrations of (a) $1:3$ sub-harmonic synchronization, and (b) $3:1$ harmonic synchronization. $t_{\mathrm{R},\mathrm{M}}$ and $t_{\mathrm{R},\mathrm{S}}$ are the roundtrip times of the cavity solitons in the master and slave resonators, respectively. CW: continuous-wave.}
\label{fig:illustrate}
\end{figure}

\begin{figure*}[t]
\centerline{\includegraphics[width=0.85\textwidth]{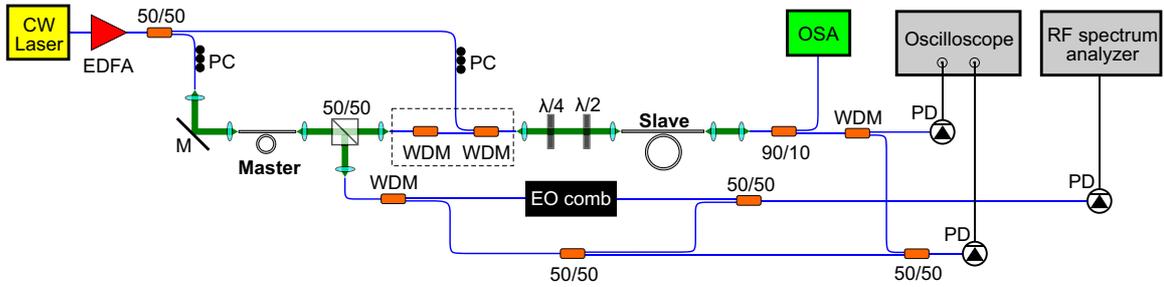}}
\caption{A schematic of our experimental setup. CW: continuous-wave, EDFA: erbium-doped fiber amplifier, PC: polarization controller, M: mirror, WDM: wavelength-division-multiplexer, $\lambda/4$: quarter-wave plate, $\lambda/2$: half-wave plate, OSA: optical spectrum analyzer, PD: photodiode, EO: electro-optic. The details on the EO comb generation can be found in the Supplemental Material~\cite{supplemental}. The green, blue and black curves correspond to the free space, fiber and electrical paths, respectively. The black dashed box indicates the coupling link.}
\label{fig:expschematic}
\end{figure*}

We schematically illustrate $1:3$ sub-harmonic synchronization and $3:1$ harmonic synchronization in Figs.~1(a) and (b), respectively. The top microresonator in each schematic serves as the master resonator and the output derived from it is used to enslave the slave microresonator below. For the sub-harmonic regime in Fig.~\ref{fig:illustrate}(a), the free-spectral-range (FSR) of the master resonator, and hence the spectral spacing of the resulting Kerr comb, is approximately thrice that of the slave resonator. As a result, each comb line from the master resonator couples to every third mode in the slave resonator. This regime can alternatively be interpreted in the time domain as synchronous pulsed driving in which the driving frequency is approximately three times larger than the natural frequency of the driven oscillator. The converse occurs for the harmonic regime in Fig.~\ref{fig:illustrate}(b), where every third comb line from the master resonator couples to the adjacent mode in the slave resonator. In the time domain, the driving frequency is lower than the natural frequency of the oscillator by a factor of three and the oscillator encounters the driving signal only once every three roundtrips.

A schematic of our experimental setup is shown in Fig.~\ref{fig:expschematic}. We employ three circular $\mathrm{Si}_3\mathrm{N}_4$ microring resonators with radii of $104.18~\mathrm{\mu m}$, $208.36~\mathrm{\mu m}$ and $312.54~\mathrm{\mu m}$, corresponding to approximate FSR's of $216$~GHz, $108$~GHz and $72$~GHz, respectively. Note that the lengths of the microresonators have been chosen so that the FSR's of the larger rings are related to that of the smallest ring by rational factors 1/2 and 1/3. The cross-sections of all our waveguides are $730\times1500$~nm. We choose two out of the three microresonators for each experiment and set them up in the master-slave configuration~\cite{jang_synchronization_2018}, as illustrated earlier in Fig.~\ref{fig:illustrate}. We examine all six possible permutations in turn. As a pump source, we use a single tunable CW laser centered at 1551.7~nm. We amplify the output of the laser with an erbium-doped fiber amplifier~(EDFA) and split it into two beams of equal power, each of which pumps one of the microresonators. The pump beams are coupled into the integrated bus waveguides with aspheric lenses, and fiber polarization controllers are used to excite the TE mode inside the waveguides. Using two narrow-band wavelength division multiplexers~(WDM's) in series, we remove the residual pump from the output of the master resonator and combine the filtered signal with the second pump. The combined field, which now encompasses the coupling signal from the master resonator, pumps the slave resonator~\cite{jang_synchronization_2018}. Overall, the length of the coupling link is approximately 1~m, including 25~cm of dispersion-compensating fiber~(DCF). It should be noted that there is nothing fundamental about choosing this length of the coupling link which is, in principle, limited only by the coherence length (approximately 2~km) of our current CW pump source~\cite{jang_synchronization_2018}. We monitor the optical spectrum of the output of the slave resonator with an optical spectrum analyzer~(OSA). We also detect the beat-note between the two comb signals on an oscilloscope, after filtering out the pump component using a WDM filter. For each microresonator, we thermally access and operate in the single-cavity soliton-based mode-locked regime~\cite{joshi_thermally_2016}, in which the comb spacing is determined by the repetition rate of the circulating cavity soliton in the microresonator. Thermo-optic tuning of microresonators enables fine sweeping of the comb spacings~\cite{bao_soliton_2017}, which allows us to examine the synchronization behavior as a function of a mismatch between the spacings of the master and slave combs. In all our measurements, we thermally tune the master resonator while leaving the slave resonator free-running. The tuning range is carefully chosen to maintain the single cavity soliton state.

Since the spectral spacings of the Kerr combs are too large to measure directly, we utilize an electro-optic~(EO) comb~\cite{vanhowe_multiwavelength_2004,okawachi_continuous_2007,delhaye_hybrid_2012,xue_thermal_2016}, details of which are presented in the Supplemental Material~\cite{supplemental}. In essence, the EO comb stage is composed of an intensity modulator and two phase modulators in series, and produces a pulse train at a 10-GHz repetition rate, which corresponds to a frequency comb with a 10-GHz spacing. We split the output of the master resonator, spectrally isolate the pump from a split beam, and use it as the input source for the EO comb. Since both the Kerr combs and EO comb share a common pump source, the spacing of the master Kerr comb can be inferred as $\Omega_\mathrm{M} = m\Omega_\mathrm{mod} \pm \Delta\Omega$, where $\Omega_\mathrm{mod}$ is the modulating frequency, $\Delta\Omega$ is the beat-note frequency between the master Kerr and EO combs measured on a RF spectrum analyzer, and $m$ is an integer. In order to determine the sign of the beat-note, we measure the frequency spacing between the pump and the $n\mathrm{th}$ line of the master Kerr comb using an OSA, and divide this spacing by $n$. As $n$ is chosen to be large ($n>20$) and the beat-note is typically larger than 1~GHz, the resolution of the OSA is sufficient to determine the sign of the beat-note. Note that in the previous experimental study of Kerr comb synchronization~\cite{jang_synchronization_2018}, only the beat-note between the two Kerr combs were measured, which inherently suffers from a sign ambiguity. The EO comb allows direct characterization of the master comb spacing, and hence resolves this ambiguity.

\begin{figure}[b]
\centerline{\includegraphics[width=1\columnwidth]{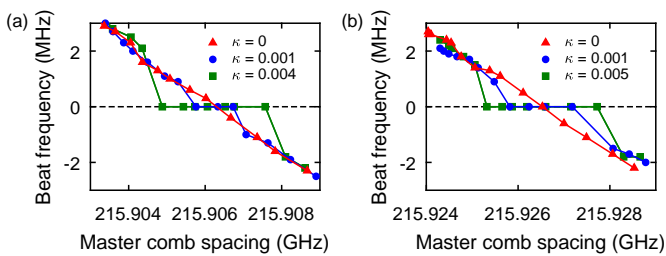}}
\caption{Experimental evidence of (a) $1:2$, and (b) $1:3$ sub-harmonic synchronizations for several coupling strengths $\kappa$. As explained in the main text, $\kappa$ is equal to the power transmission coefficient of the coupling link. Comb spacing of the master resonator is thermally tuned while the slave resonator is left free-running. }
\label{fig:subharmonic}
\end{figure}

Figures~\ref{fig:subharmonic}(a) and (b) show experimental evidence of $1:2$ and $1:3$ sub-harmonic synchronization, respectively. The vertical axes denote the measured beat-note frequency between the master and slave combs. The coupling strength $\kappa$ is a direct measure of the power transmission of the coupling link, that is, fraction of the master comb coupled to the slave comb. Without coupling (red curves), the beat-note changes monotonically and crosses zero at a single point as the master comb is thermally tuned. In comparison, when coupling is introduced (blue and green curves), there is a range of comb spacings where no beat-note is detected. We identify this range as the synchronization region in which the two combs can remain synchronized despite a mismatch in their mode spacings. This range increases with increasing coupling strength, which is consistent with classical theory of synchronization~\cite{pikovsky_synchronization_2003}.

\begin{figure}[b]
\centerline{\includegraphics[width=1\columnwidth]{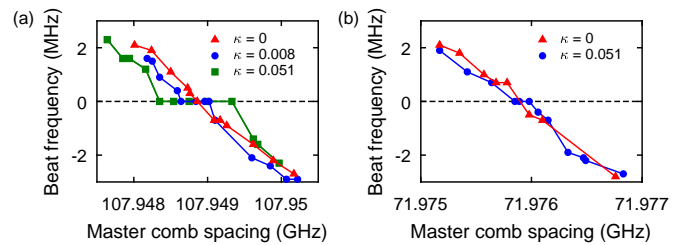}}
\caption{Experimental evidence of (a) $2:1$, and (b) $3:1$ harmonic synchronizations for several coupling strengths $\kappa$.}
\label{fig:harmonic}
\end{figure}

The regimes of $2:1$ and $3:1$ harmonic synchronization are studied in a similar manner, as shown in Fig.~\ref{fig:harmonic}. While the primary features of sub-harmonic and harmonic synchronization are similar, we notice that the synchronization ranges are significantly narrower for the latter. This aspect is particularly prominent in Fig.~\ref{fig:harmonic}(b), where the synchronization range is very narrow even at the largest coupling strength accessible experimentally, $\kappa = 0.051$ (limited by losses in the coupling link), and becomes progressively narrower as the coupling strength is reduced (results at lower coupling strengths are omitted for visual clarity). Such a dramatic difference in the synchronization range can be understood from the fact that for sub-harmonic synchronization, the driving frequency is higher than the comb spacing of the slave resonator by an integer factor [factor of three in Fig.~\ref{fig:illustrate}(a) and Fig.~\ref{fig:subharmonic}(b)]. The cavity soliton in the master resonator completes three roundtrips in its host resonator per one roundtrip of the cavity soliton in the slave resonator and as a result, the latter encounters the coupling signal every roundtrip. However, for the harmonic case in Fig.~\ref{fig:illustrate}(b) and Fig.~\ref{fig:harmonic}(b), the driving frequency is lower by a factor of three and the soliton in the slave resonator is influenced by the coupling signal only once per three roundtrips.

\begin{figure*}[t]
\centerline{\includegraphics[width=1\textwidth]{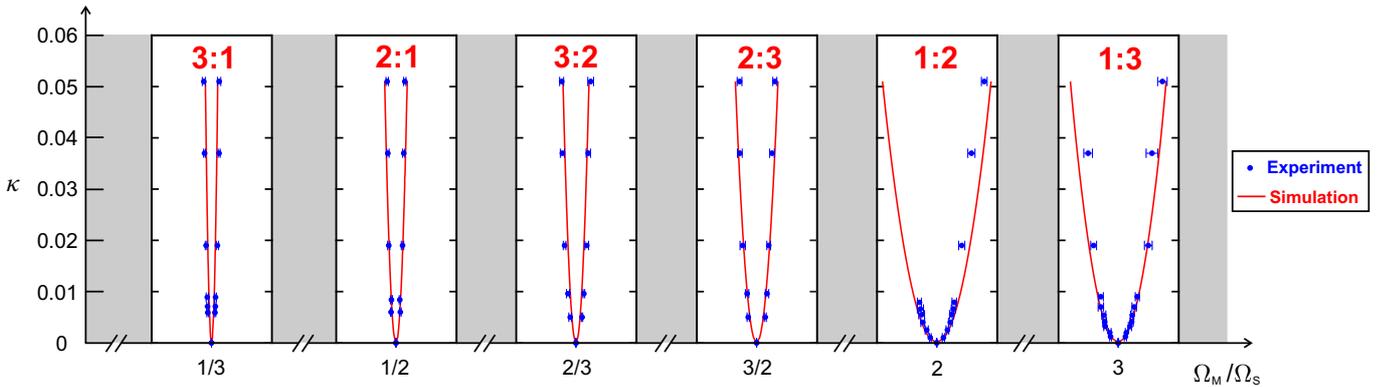}}
\caption{Arnold tongues for different regimes of synchronization. The blue dots represent experimental results while the red curves on the same axes correspond to our numerical simulations. Each experimental data point results from statistical analysis of 20 repeated measurements. Error bars indicate 2 standard errors. All white regions correspond to an absolute frequency range from $-7$~MHz to $7$~MHz of the mismatch $\delta f = r\Omega_{\mathrm{S}}-q\Omega_{\mathrm{M}}$. Grey shades indicate regions that are not investigated. Axis breaks are used on the x-axis for clearer representation of the results.}
\label{fig:arnold}
\end{figure*}

The plots in Figs.~\ref{fig:subharmonic} and \ref{fig:harmonic} allow us to estimate tolerable frequency mismatches between the master and slave combs for several coupling strengths. We examine these features in more detail for various orders of synchronization and present our results in Fig.~\ref{fig:arnold}. Experimental data are plotted as blue dots while the corresponding numerical results based on coupled Lugiato-Lefever equations~\cite{jang_synchronization_2018,lugiato_spatial_1987,daguanno_nonlinear_2016,daguanno_coupled_2017} are superimposed as red curves on the same axes (see Supplemental Material~\cite{supplemental} for more details on numerical simulation). Each white window corresponds to an absolute frequency range from $- 7$ to $7$~MHz of the mismatch $\delta f = r\Omega_{\mathrm{S}}-q\Omega_{\mathrm{M}}$, while the grey shades are used to denote regions that are not investigated in our study. We also use axis breaks to omit the irrelevant regions for clearer visual presentation of our results. Since the optical path length of the 1-m-long coupling link in our setup is not stabilized, the phase of the coupling signal is susceptible to a slow drift over the course of our measurements. In order to account for this drift, we performed 20 repeated measurements for a given $\kappa$. Each experimental point results from statistical analysis of the 20 measurements, where the error bar denotes two standard errors. We also numerically emulate our experimental conditions by carrying out simulations for a range of coupling phases covering the entire range of $2\pi$, in steps of $2\pi/36$, and averaging the results. Our numerical analysis is in good agreement with the experiment. The curves of Fig.~\ref{fig:arnold} are equivalent to Arnold tongues in the theory of dynamical systems~\cite{pikovsky_synchronization_2003}. We also note that rational, non-harmonic orders of synchronization such as $2:3$ and $3:2$ regimes are possible, which is consistent with the extensive scientific literature of synchronization~\cite{pikovsky_synchronization_2003}. We attribute slight asymmetry in the Arnold tongues of Fig.~\ref{fig:arnold} to the asymmetric frequency-dependence of the power transmission coefficient in the coupling link (see Fig.~S5 in the Supplemental Material~\cite{supplemental}). We have numerically confirmed that removing this frequency-dependence restores the symmetry of the structures.

With the current setup, we are unable to obtain measurements for the lower bounds of the synchronization regions at larger $\kappa$ for $1:2$, and at $\kappa = 0.051$ for $1:3$ sub-harmonic synchronizations. This is due to a technical difficulty in simultaneously aligning the center of the synchronization region (where the ratio of the comb spacings is exactly a rational number) with the two centers of the detuning ranges of stable comb operation in the two coupled microresonators. This complication is compounded by the fact that the synchronization regions of the sub-harmonic regimes are very broad for larger coupling strengths. On the other hand, the synchronization regions of other regimes are much narrower and such issues do not impede our measurements. We note slight deviations between the experimental and numerical results in Fig.~\ref{fig:arnold} which are more prominent for the sub-harmonic regimes at larger coupling strengths. We believe these deviations are a consequence of coupling. More specifically, near the vicinity of the synchronization transition, the relative drift rate of the cavity soliton in the slave resonator decreases locally as it encounter the coupling signal. As a result, the experimentally observed (effective) beat-note frequency is slightly less than that expected in the absence of coupling. Another potential source of deviation is the nonlinear response of the integrated heater. The spacing of the master Kerr comb changes approximately in proportion to the power dissipated by the integrated heater, which in turn varies quadratically with the voltage across it. The DC voltage across the heater thus determines the rate of variation of the comb spacing per unit voltage, or equivalently, the resolution of our measurement. In order to simultaneously excite combs in both microresonators at a common pump wavelength, high DC voltages ($> 6$~V) were often employed which could have resulted in overestimation of the ranges of synchronization regions for the $3:1$ and $3:2$ regimes.

Our study presents the first experimental and theoretical results on sub-harmonic, harmonic and harmonic-ratio synchronization in systems of coupled cavity soliton Kerr combs. Each distinct synchronization regime of order $q:r$ (corresponding to a distinct ratio of $q$ to $r$) is observed to exhibit an Arnold tongue structure as the coupling strength is varied. The appearance of Arnold tongues is ubiquitous in nonlinear dynamical systems, which demonstrates the general applicability of our results. In principle, our system can be extended to a larger number of oscillators on an integrated platform, which promises a novel compact platform for the implementation and emulation of complex networks under a well-controlled environment. In addition, our scheme can be applied to passively lock the frequencies of Kerr combs whose spacings are related by a rational factor. It becomes especially useful when one of the synchronized combs has an electronically detectable spectral spacing. Our technique can facilitate the characterization and stabilization of a widely spaced Kerr comb in a way similar to the EO comb-based technique~\cite{delhaye_hybrid_2012}. We believe that a sub-harmonic regime of synchronization is more suitable for practical applications. Although there is no conceivable physical limitation in the harmonic ratios that can be achieved for neither sub-harmonic nor harmonic regimes, Fig.~\ref{fig:arnold} and our physical intuition imply that the synchronization region of a harmonic regime would shrink rapidly as the harmonic ratio becomes larger. Finally, we emphasize that our coupling scheme is linear~\cite{jang_synchronization_2018} and independent of the details of system nonlinearity. It could therefore broadly impact applications that demand fully stabilized Kerr combs, such as precision spectroscopy and metrology.

\subsection*{}
$^{*}$Corresponding author: alg2207@columbia.edu

\newcommand{\enquote}[1]{#1}

\bigskip

\section*{Acknowledgements}
\noindent The authors acknowledge supports from Air Force Office of Scientific Research (AFOSR) (grant no. FA9550-15-1-0303), National Science Foundation (NSF) (grant no.
CCF-1640108), Semiconductor Research Corporation (SRC) (grant no. SRS 2016-EP-2693-A), and Lockheed Martin Advanced Technology Center.

%\section*{Additional information}

\end{document}